\documentclass[aps,floats,epsf,twocolumn]{revtex4}
 \usepackage{graphics}

\begin{document}

\title{Stable Skyrmions in spinor condensates}

\author{ Igor F. Herbut$^{1,2}$ and Masaki Oshikawa$^{2}$ }

\affiliation{$^1$ Department of Physics, Simon Fraser University,
 Burnaby, British Columbia, Canada V5A 1S6\\ $^2$
 Department of Physics, Tokyo Institute of Technology, Oh-okayama,
 Meguro-ku, Tokyo 152-8551, Japan}

\begin{abstract}
Globally symmetric spinor condensates in free space are
argued not to support stable topological defects in either two or
three dimensions. In the latter case, however, we show that a  topological
Skyrmion can be stabilized by forcing it to adopt certain density
profiles. A sufficient condition for the existence of
Skyrmion solutions in three dimensions  is formulated and illustrated in
simple examples. Our results pertain to Bose-Einstein condensation
in $^{87} Rb$.
\end{abstract}
\maketitle

\vspace{10pt}

Topologically non-trivial configurations play a crucial role in our
present understanding of ordering, dynamics, and criticality in
basic models of statistical mechanics \cite{chaikin}. Probably the best understood
example is that of vortices and vortex-loops in the Ginzburg-Landau, or
Higgs, $O(2)$-symmetric theory for a complex order parameter in two
(2D) and three dimensions (3D), which are responsible for the very
existence of the disordered phase at high temperatures
\cite{herbutbook}. The next in order of increasing complexity is the
$O(3)$-symmetric  Heisenberg model, which allows texture-like
Skyrmions in 2D \cite{belavin}, and point-like hedgehogs in
3D \cite{manton}. Whereas Skyrmions in 2D are
understood to provide an additional source of  disorder for the
low-temperature phase, the exact role of hedgehogs in the 3D Heisenberg
model is already less clear \cite{dasgupta}. Both, nevertheless, represent
non-trivial spatial configurations, which by virtue of their
distinct topology are locally stable and separated from the ground
state by an infinite energy barrier.

In this Letter we study the issue of topological defects in the next
simplest case: the order parameter with global $O(4)$ symmetry.
Such a symmetry arises in complex condensates with an internal
spin-$1/2$-like quantum number \cite{lieb}, for example.
Realizations of such {\it spinor
condensates} are found in models of inflatory cosmology
\cite{vilenkin}, Bose-Einstein condensation of $^{87}Rb$
\cite{ueda}, bosonic ferromagnetism \cite{lieb}, \cite{saiga}, and in
effective theories of high-temperature superconductivity \cite{lee}, \cite{herbut},
 and of deconfined criticality \cite{senthil}.  The Higgs sector of the
 Weinberg-Salam model of electro-weak interactions
 represents another closely related example, with a spinor condensate coupled
 to gauge fields. It is easy to show that the triviality of
the first and the second homotopy groups of the three-dimensional
unit sphere, $S_3$, on which the order parameter lives at low
temperatures, implies that there are no stable topological defects
in two dimensions. 2D spinor condensates may therefore
be expected to be in the disordered phase at all finite temperatures,
precisely as described by the $O(4)$ non-linear $\sigma$-model ($NL
\sigma M$). In 3D, on the other hand, standard topological
considerations suggest the possibility of a Skyrmion texture,
thanks to the {\it third} homotopy group of $S_3$ being
the group of integers. Here, however, unexpected subtleties arise,
which together with the growing relevance of the
problem provide the motivation for the present work.

It has been known that finite-energy textures in higher than two dimensions are generally
unstable with respect to shrinkage \cite{hobart}.
Derrick's scaling argument, does not, however, forbid topological defects
with infinite energy, hedgehogs in the 3D Heisenberg model being a prime
example. The search for stable topological defects in the $O(4)$ $NL\sigma
M$ in 3D, however, has so far led to a negative result
\cite{durrer}, \cite{stoof}, \cite{anglin}, \cite{battay}. We first demonstrate that
even with amplitude variations included the spherically symmetric
Skyrmion remains unstable. This conclusion follows from a useful
mechanical interpretation of Skyrmion's differential equation,
in which the radial dependence of particle's density appears
as a `source of dissipation' for a fictitious classical
particle, which prevents it from oscillating with a full amplitude.
This hindrance, however, may be turned into an
advantage by {\it forcing} the same radial dependence to have a form that
effectively serves as an `energy pump'.
This idea yields an integral condition on the density profile of a
Skyrmion, which we illustrate with three qualitatively different analytical solutions
of the non-linear Euler-Lagrange's equations. Skyrmion solutions are found, for example,
for specific forms of the confining potential, which may  be used
in creating such configurations in the laboratory \cite{ueda}.

Let us begin by  defining the system of two-component
 bosons in the continuum  with the standard action in terms of
 the complex coherent states written as
 \begin{equation}
 S=\int_0 ^\beta d\tau d^3 \vec{x} [  \Phi^\dagger
 (\partial_\tau - \nabla^2 -\mu +V(\vec{x}) ) \Phi +
 \frac{U}{2}
 (\Phi^\dagger \Phi)^2 ],
 \end{equation}
where $\Phi^\dagger = (\Phi_1 ^* (\vec{x},\tau), \Phi_2 ^*
(\vec{x},\tau))$.
We have also included a confining potential $V(\vec{x})$,
assumed a contact repulsion $U$ for simplicity, and set $\hbar=1$, and the boson
mass $2m=1$.  $\beta=1/k_B T$ is the inverse temperature.

We are interested in finding classical, that is
$\tau$-independent, locally stable field configurations
of the action (1). The field $\Phi(\vec{x})$ may be written
in terms of its real amplitude $f(\vec{x})$ and the normalized
complex spinor $a (\vec{x})$ as $\Phi^\dagger
= f a^\dagger$, with $a^\dagger a =1$. The low-energy physics
is then described by fluctuations of the spinor $a$, which can be identified
with a four-component real vector of unit length. The symmetry of the order
parameter is thus $O(4)$.  Varying the above action
leads to the requisite Euler-Lagrange's equations
\begin{equation}
f( a^\dagger \nabla^2 a+(\nabla^2 a^\dagger) a ) a = 2 f \nabla^2 a + 4 (\nabla f) \cdot
(\nabla a),
\end{equation}
\begin{equation}
-\nabla^2 f + [ (\nabla a^\dagger)\cdot (\nabla a) - 1 + V] f + f^3
=0,
\end{equation}
where we have rescaled the lengths as $\vec{x} \sqrt{\mu}
\rightarrow \vec{x}$, the amplitude as $f \sqrt{U/\mu}\rightarrow f$,
and the external potential as $V/\mu \rightarrow V$. Solution
needs to satisfy the boundary conditions
\begin{equation}
 \lim_{x\rightarrow \infty} x^2 f^2 \nabla a  =0,
\end{equation}
 which guarantee stability with respect
 to small rotations of the spinor $a$ at the infinitely
remote boundary of the system.

Eqs. (2) and (3) form a set of {\it five} differential equations for
real and imaginary parts of the spinor $a$ and the amplitude
$f$. In a spherically symmetric potential $V(\vec{x})$
we can partially solve them, however, by assuming the most general
{\it ansatz} with the same symmetry \cite{manton}:
$f(\vec{x})=f(r)$,
\begin{equation}
a^\dagger (\vec{x}) = ( \sin \omega(r) \cos\theta + i \cos
\omega(r), \sin \omega(r) \sin \theta e^{i\phi}),
\end{equation}
where  $\vec{x} = (r \sin\theta \cos\phi, r \sin\theta \sin \phi,
r\cos\theta)$. Somewhat tedious but otherwise straightforward
 algebra shows that {\it all four} of
Eqs. (2) will then be satisfied provided that  the function
$\omega(r)$ satisfies the differential equation
\begin{equation}
\frac{d^2  \omega(r)}{dr^2} + [ \frac{2}{r}+
\frac{2}{f(r)}\frac{df(r)}{dr} ] \frac{d \omega(r)}{dr} - \frac{\sin
(2 \omega(r))}{r^2} =0.
\end{equation}
Similarly,  Eq. (3) then reduces to
\begin{eqnarray}
-\frac{d^2 f(r)}{dr^2}-\frac{2}{r}\frac{df(r)}{dr} +[
(\frac{d\omega(r)}{dr})^2 + \\ \nonumber 2 (\frac{\sin
\omega(r)}{r})^2 -1 +V(r)] f(r) + f^3 (r) =0.
\end{eqnarray}
The action for the ansatz (5) is
\begin{eqnarray}
\frac{S}{4\pi \beta} = \int_0 ^\infty  r^2 [ (\frac{df}{dr})^2  + ( (\frac{d\omega}{dr})^2
+\\ \nonumber
2(\frac{\sin\omega}{r})^2 +V(r) -1)f^2  + \frac{f^4}{2} ]dr,
\end{eqnarray}
so that Eqs. (6) and (7) may also be recognized as Euler-Lagrange's equations
for the latter form of $S$.

The  ansatz for $a(\vec{x})$ will be topologically non-trivial
and wrap the $S_3$, defined by the condition $a^\dagger a=1$, $N$ times if, as the
radius varies from zero to infinity, the function $\omega(r)$ takes
all the values from zero to $N\pi$. Hereafter we restrict our
discussion only to the elementary Skyrmion with $N=1$.

If we turn the external potential off ($V(r)\equiv 0$), it is easy to show
that the Eqs. (6) and (7) {\it do not} actually have the Skyrmion as a
solution, the assumptions to the contrary notwithstanding \cite{stoof}, \cite{anglin}.
To this purpose define a new `time' variable $-\infty < t=\ln r < \infty$.
Eq. (6) then becomes
\begin{equation}
\ddot{\omega}(t) = - \frac{d W(\omega)}{d\omega} - \eta(t)
\dot{\omega}(t),
\end{equation}
with $W(\omega)=(\cos 2 \omega)/2$, $ \eta(t)= 1+ d\ln f^2 (t) /dt$,
and $\dot{\omega}=d\omega/dt$. The function
$\omega(t)$ may now be interpreted as
a coordinate of a particle moving in the potential $W(\omega)$,
coupled to a dissipative environment with the {\it time-dependent}
coefficient of dissipation $\eta(t)$. Let us first neglect the
amplitude variations dictated by Eq. (7).
Dissipation then still persists, and $\eta(t)\equiv 1$. Demanding that
$\omega(0)=0$ Eq. (6) implies $\omega(r)\sim r$ for $r\ll 1$, so that the
initial condition in our mechanical analogy is $\omega(-\infty)=
\dot{\omega}(-\infty) = 0$. This corresponds to rolling down from
the top of the potential $W(\omega)$ without any initial kinetic energy.
Independently of the slope of $\omega(r)$ at the origin the solution therefore
always oscillates around and approaches $\omega(\infty)=\pi/2$, dissipating energy and
loosing its amplitude in the process. Eq. (6) for a fixed amplitude therefore
admits only a {\it meron} (half-Skyrmion) as a solution. However, since when
$r \gg 1$, $|\omega(r)-\pi/2| \sim 1/\sqrt{r}$ for the meron, the boundary
condition in Eq. (4) is not satisfied. Consequently, the meron
 is {\it unstable} with respect to deformation $\omega(r)
 \rightarrow \omega(r) - \epsilon (r)$ with $\epsilon (0)=0 $
 and $0< \epsilon(\infty)\ll 1$ \cite{remark}.

Before relaxing the condition of fixed amplitude let us observe that
solutions of Eq. (6) (still with $f(r)=const$) that  satisfy the boundary conditions (4) and
approach $\omega(\infty)=\pi$ as $\omega(r) \sim \pi \pm  1/r^2$, do in fact exist. These solutions, however,
are singular  near the origin, $\omega(r) \sim \pm 1/r$, and
hence have an infinite action. We will refer to these as
{\it singular} solutions and discuss their possible significance in a moment.

\begin{figure}[t]
{\centering\resizebox*{80mm}{!}{\includegraphics{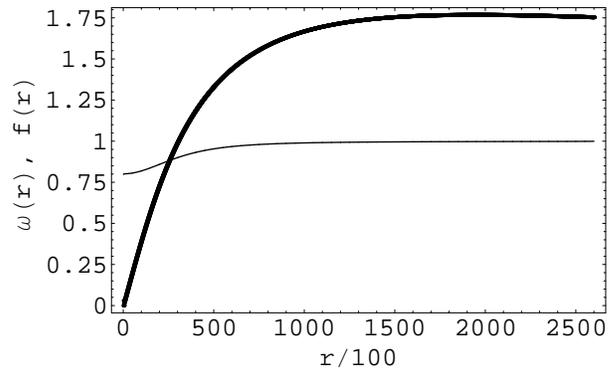}} \par}
\caption[] { A numerical solution of Eqs. (6) and (7): $\omega(r)$ (thicker curve) and $f(r)$.}
\end{figure}

Next, we retain $V(r)\equiv 0$, but allow the amplitude
variations. Finiteness of the action, measured from the
value for the
ground state $f(r)\equiv 1$, $\omega(r)=0$, requires then $f(\infty)=1$.
Linearizing near the origin one still finds $\omega(r)=A r +O(r^2) $,
and $f(r) = f(0)+ (f(0)/6) (f^2 (0) + 3 A^2 -1) r^2 +O(r^3)$, where
$A$ and $f(0)$ are finite constants. To have $f(\infty)=1$ we then
find numerically that $f(0)<1$, and $f(r)$ to be a weakly and
uniformly increasing function (Figure 1). The reason behind such a behavior is
obvious: gradients of $a(\vec{x})$ are largest near the origin,
so suppressing the amplitude there somewhat always lowers the energy. Energy density
being integrable, however, the amplitude remains finite at the origin, and
the Skyrmion is `coreless'.
With such a solution for $f(r)$, however, one still finds $\omega(\infty)
  =\pi/2$ and Eq. (4) violated. This is evident from
  our mechanical analogy in Eq. (9)
  in which having the amplitude increasing with radius only
  adds to `dissipation'.

The stage is now set for the main part of the Letter.
The preceding analysis points to a simple way to
stabilize the Skyrmion. To have a solution of Eqs. (6) and (7)
satisfying the requisite
boundary conditions $\omega(t=-\infty) = \dot{\omega}
(t=-\infty)=0$, and $\omega(t=\infty)=\pi$, $\dot{\omega}(t=\infty)=
0$, it is necessary and sufficient that the total dissipation
 in our mechanical analogy vanishes. Mathematically \cite{landau}:
\begin{equation}
\int_{-\infty}^{\infty}  \eta(t) \dot{\omega}^2(t) dt=0.
\end{equation}
The total mechanical, and thus the potential,
 energy is then the same at the initial and the final `time' $t$.
 The amplitude therefore must be a {\it decreasing}
function at least for some radia, where it would `pump' the energy back
into the oscillator. This is our main result.
In the remaining we provide some
specific amplitude profiles that indeed lead to stable Skyrmions.

1) The trivial case is the one without dissipation:
$\eta(t)\equiv 0$. This is equivalent to the particle density
\begin{equation}
f^2 (r) = \frac{A}{r},
\end{equation}
with $A$ as a constant. This yields
\begin{equation}
\omega(r) = 2 \cot^{-1} (r^{\sqrt{2}}).
\end{equation}
Inserting this solution into Eq. (7) gives the external potential
that enforces such a density to be
\begin{equation}
V(r) = 1-\frac{A}{r} - \frac{1}{4 r^2} - \frac{ 16
r^{2(\sqrt{2}-1)}}{(1+r^{2\sqrt{2}})^2 }.
\end{equation}
It is easy to check that this solution has a finite action,
due to the singularity of the external potential at the origin.
The same singularity, however, makes this simplest example
less than completely satisfying from a possible
 practical point of view.

2) The difficulty in utilizing the condition in Eq. (10) is
that the velocity $\dot{\omega}(t)$ of course depends on the dissipation
$\eta(t)$ in a rather non-trivial way, so it appears that solving  the
differential equations (6) and (7) is always unavoidable in practice.
This, fortunately, is not so, due to
 the following theorem: {\it For potentials $W(\omega)$
that are even functions of $\omega-\pi/2$, and dissipation
coefficients $\eta(t)$ that are odd functions of $t$, solutions of
Eq. (9) are such that}
\begin{equation}
\omega(t)-\frac{\pi}{2} = - \omega(-t) + \frac{\pi}{2},
\end{equation}
i.e. odd functions of time, measured from the moment when at
the bottom of the potential. It is thus sufficient
to have a density profile that implies an odd $\eta(t)$
to stabilize the Skyrmion. As an illustration, consider
\begin{equation}
\eta(t)= - \tanh (t),
\end{equation}
for which the the particle density is simply
\begin{equation}
f^2 (r)=\frac{B}{1+r^2},
\end{equation}
with $B$ as a constant.
It is straightforward to check that the solution of Eq. (7) is then
\begin{equation}
\omega(r) = 2 \cot^{-1} (r),
\end{equation}
and the corresponding external potential
\begin{equation}
V(r)= 1 -\frac{B}{1+r^2}- \frac{15}{(1+r^2)^2}.
\end{equation}
Different density profiles can be similarly constructed, including
those that would lead to a finite total number of particles.

We may  note that for the external potential in
Eq. (18) there is yet another, topologically trivial,
 solution of Eqs. (6) and (7):
$\omega(r)=0$, $f_0 (r)$. Numerical solution for $f_0 (r)$ for $B=1$
is plotted on Figure 2. Since $f_0 (r) > f(r)$, the action, which at a stationary
point takes the form
\begin{equation}
\frac{S}{4\pi \beta} = -\frac{1}{2} \int_0 ^\infty r^2 f^4 (r) dr,
\end{equation}
is actually lower for $f_0 (r)$. This represents the
ground state configuration in the potential (18).

\begin{figure}[t]
{\centering\resizebox*{80mm}{!}{\includegraphics{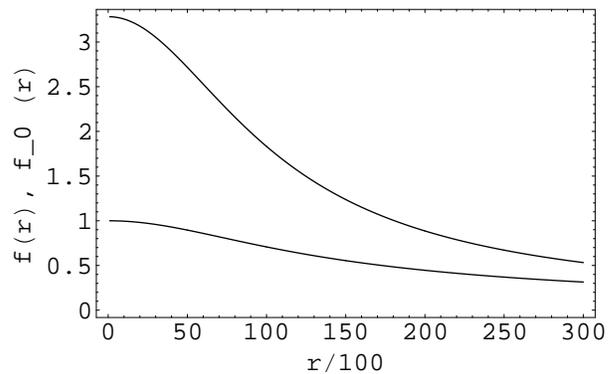}} \par}
\caption[] { Topologically trivial ground state $f_0 (r)$    (upper curve) and the Skyrmion's amplitude
$f(r)$, in the potential in Eq. (18) with $B=1$. }
\end{figure}

3) Finally, the Skyrmion may be stable in a constant, but discontinuous,
 particle density as well. Take $f^2(r)=e^C$ for $r<1$, and $f^2 (r) = 1$ for $r>1$,
for example. Then
\begin{equation}
\frac{d\ln f^2 (t)}{dt} = -C \delta(t),
\end{equation}
and the total dissipation will vanish for
$C$ tuned to
\begin{equation}
C= \int_{-\infty} ^{\infty} [\frac{\dot{\omega}(t)}{
\dot{\omega}(0)}]^2  dt.
\end{equation}
Such a discontinuity in  the amplitude may be understood as providing a
`kick' in our mechanical terminology, which instantaneously increases the
kinetic energy at $t=0$ so that there is just enough to reach the
next top of the potential $W(\omega)$ at $\omega=\pi$. This is possible because for any
regular solution of Eq. (6) (with $f=const$) which starts at $\omega(r=0)=0$ there is a
singular solution that intersects it at $r=1$, and  which asymptotically
approaches $\pi$ as the radius increases. Their difference in slopes at $r=1$
determines the required injection of energy measured by the discontinuity $C$.

  Previously, the Skyrmion was found  to be stable numerically if the
  $O(4)$ symmetry of the action in Eq. (1) is broken down to
  $O(2)\times O(2)$, in a way that favors phase separation \cite{battay}.
  In contrast, our action is fully $O(4)$ symmetric, and the
  stability obtains from the interplay between
  the form of the Skyrmion and its density profile. Derrick's
  scaling argument is likewise evaded by the introduction of the preferred
  length scale for density variation. We hope our results
  will  help facilitate the production of these interesting
  topological objects in laboratory spinor condensates, such as
  $^{87}Rb$. Recent advances in (dynamic) manipulation of confining
  potentials seem particularly promising in this respect \cite{boyer}.

Finally, for completeness let us revisit the issue of defect's
stability in 2D. Consider
\begin{equation}
a^\dagger (\vec{x}) =
( \sin \lambda \cos\theta(r)  + i \cos
\lambda, \sin \lambda  \sin \theta(r) e^{i\phi}),
\end{equation}
 where $\vec{x}= r(\cos\phi, \sin\phi)$, and $\theta(0)=0$ and
 $\theta(\infty)=\pi/2$. For a fixed parameter $\lambda=\pi/2$ this reduces to the
configuration often discussed in literature \cite{ueda}.
It is evident, however, that by tuning $\lambda$  one may
continuously deform the defect at $\lambda=\pi/2 $ to the trivial
vacuum at $\lambda=0$, monotonically decreasing the action along the way.
This construction exploits the fact that $\pi_2 (S_3)=1$.
Topological defects in spinor condensates are thus in general
unstable in free space both in 2D and in 3D, although for
rather different reasons.

This work has been supported by NSERC of Canada (IFH), Grant-in-Aid for
 Scientific Research (MO), and 21st Century COE Program at Tokyo Institute
 of Technology, `Nanometer scale Quantum Physics' (IFH and MO), from MEXT of Japan.
 The authors also acknowledge helpful conversations with K. Ino, M. Kennett,
 J. McGuirk, B. Seradjeh, and M. Ueda.

\end{document}